\documentclass{article}
\usepackage{graphicx} 
\usepackage[final]{hyperref} 
\usepackage[margin=.75in,letterpaper]{geometry} 
\usepackage{siunitx} 
\usepackage{subcaption} 
\usepackage{authblk} 
\usepackage{multirow}
\providecommand{\keywords}[1]
{
  \small	
  \textbf{\textit{Keywords---}} #1
} 
\hypersetup{
	colorlinks=true, 
	linkcolor=blue, 
	citecolor=blue,
	filecolor=magenta,
	urlcolor=blue         
} 

\begin{document}

\title{Oignon: Citation Graph Tool}
\author{H. Ballington}
\affil{Institute for Atmospheric and Environmental Research \\
University of Wuppertal \\
Germany}
\date{\today}
\maketitle
\begin{abstract}
Citation graph visualisation is a useful tool for contextual awareness in academic research. Unfortunately, existing solutions can suffer from several drawbacks, such as a poor scaling, shallow network traversal, freemium gating, and slow build times. Oignon is a free, open-source tool for systematically exploring academic research. It uses a dual-path ranking system with recency weighting to create graphs capturing both foundational works and recent breakthroughs related to a specific publication.
\end{abstract}

\begin{figure}[hbtp]
    \centering
    \includegraphics[width=0.5\linewidth]{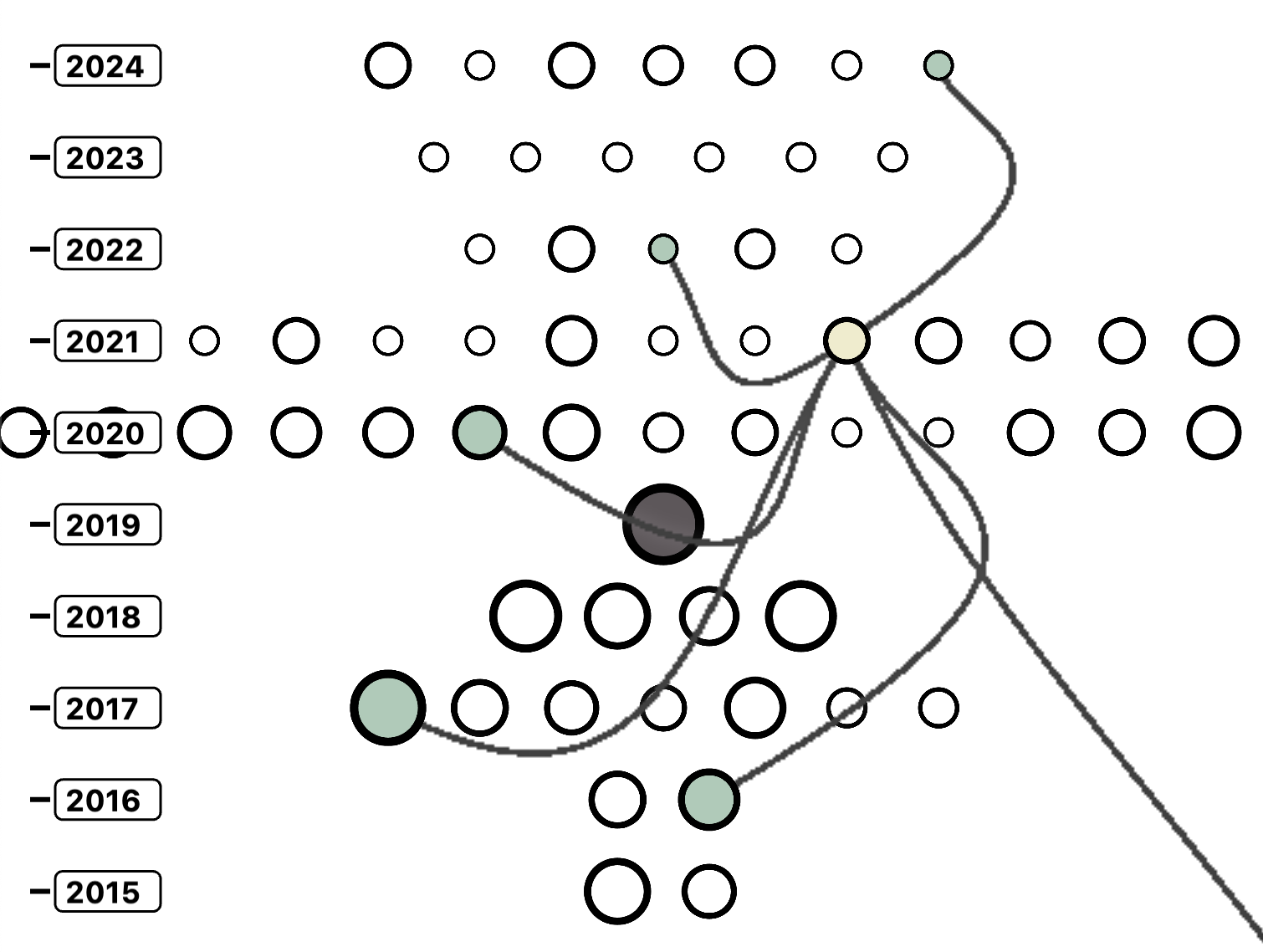}
    \caption{Example subsection of a graph created with Oignon.}
    \label{fig:main}
\end{figure}


\section{Background}

There are \textit{many} existing tools for creating citation graph visualisations, but almost all of them suffer from at least one of the following drawbacks (in order of nuisance):
 
\begin{enumerate}
    \item \textbf{Barrier to entry:} Despite the free availability of accessible data like OpenAlex~\cite{OpenAlex} and Semantic Scholar~\cite{SemanticScholar}, many citation graphing tools impose artificial barriers: mandatory account creation, freemium feature gating, or institutional authentication.
    \item \textbf{Force-directed graphs:} Most modern tools display graphs as force-directed graphs, where nodes representing papers are positioned via an iterative physics simulation. Nodes repel while connections attract, which naturally leads to a layout that prioritises topological structure over semantic meaning: a paper's position conveys little beyond its immediate neighbourhood. Such layouts also suffer from an $O(n^2)$ visual complexity - edge crossings and node occlusions grow quadratically with graph size, leading to a 'hairball effect' scalability issue that makes the graph difficult to read in a systematic way.
    \item \textbf{Poor forward-looking coverage:} Many tools focus on backward traversal, producing archival snapshots frozen at the publication date. Forward coverage tends to systematically favour older works that have had longer to accumulate citations. Temporal normalisation in this area is an important factor that needs to be accounted for. In addition, backward traversal is inherently easier to deal with because it benefits from a convergent structure - references tend to funnel toward shared foundational works, amplifying the co-citation signal. Conversely, forward traversal has a divergent structure, where citing papers branch into a variety of research avenues with weak co-citation signals; this often leads to slow graph construction that most tools either ignore or solve by truncating the forward scope.
    \item \textbf{Depth \& completeness:} The rigour of academic writing is correlated with an understanding of the work within a broader intellectual lineage. Identification and acknowledgement of foundational works and pioneering papers is an essential facet of scholarly depth. Existing tools largely fail to surmount this knowledge that has traditionally been tacit expertise, usually because the citation network traversal is shallow, which fails to uncover the deeper roots that the references have in common. 
\end{enumerate}

\section{Oignon}
Oignon (\textit{"onion" in French}) is a tool for creating citation graphs of academic research. The source code is available under the MIT license~\cite{Oignon}. Nodes in a graph are rendered on a 2D grid, with the vertical axis used to represent the year of publication. Graphs can be created either from a source paper, or by author. When building from a source paper, Oignon uses OpenAlex to search for relevant related root (previous) and branch (future) publications using a ranking system (inspired by \cite{LocalCitationNetwork}). 
\begin{itemize}
    \item \textbf{Root Ranking:} The root seed papers are defined as the direct references of the source. The rank of root papers is a simple sum of the following metrics:
    \begin{itemize}
        \item \textit{Cocited count:} How many times the root candidate is cited alongside a seed paper (relevance)
        \item \textit{Cociting count:} How many references this paper shares with the seeds (similarity)
        \item \textit{Cited count:} How many seed papers cite the root candidate (importance)
    \end{itemize}
    \item \textbf{Branch Ranking:} The branch seed papers are defined as the publications which directly reference the source. The rank of branch papers is a simple sum of the following metrics:
    \begin{itemize}
        \item \textit{Citing count:} How many seed papers are cited by the branch candidate (relevance)
        \item \textit{Cociting count:} How many references this paper shares with the source (similarity)
        \item \textit{Cocited count:} How often this paper is cited alongside the source paper, weighted by recency (relevance + importance)
    \end{itemize}
    The recency weighting is a logarithmic weighting: $w(t) = 1 + \ln\left(1 + \frac{h}{t}\right)$, where $t$ is years since publication (clamped to~1) and $h=4$ is half-life.
\end{itemize}
\begin{table}[h]
\centering
\renewcommand{\arraystretch}{1.25}
\begin{tabular}{l l l l}
\hline
\textbf{Direction} & \textbf{Metric} & \textbf{Definition} & \textbf{Interpretation} \\
\hline
\multirow{4}{*}{\textbf{Root} (backward)} 
  & \textit{Cited count} & Seeds citing this paper & Importance \\
  & \textit{Cocited count} & Co-appearances with seeds in bibliographies & Relevance \\
  & \textit{Cociting count} & Shared references with seeds & Similarity \\
  & \textbf{Total rank} & Sum of above & Overall Score \\
\hline
\multirow{4}{*}{\textbf{Branch} (forward)} 
  & \textit{Citing count} & Seeds cited by this paper & Relevance \\
  & \textit{Cociting count} & Shared references with source & Similarity \\
  & \textit{Cocited count} & Co-citations with source \textbf{(recency-weighted)} & Relevance + Importance \\
  & \textbf{Total rank} & Sum of above & Overall Score \\
\hline
\end{tabular}
\caption{Ranking metrics for root and branch paper selection.}
\label{tab:ranking-metrics}
\end{table}
A summary of the ranking metrics is shown in Table~\ref{tab:ranking-metrics}. The final nodes selected for the graph are the source, all direct references of the source, the branch seeds (capped), and the top-ranked roots and branches. A subsection of a graph created with Oignon is shown in Figure~\ref{fig:main}. Publications are shown as circles, sorted by year of publication. The size of the circle represents the number of global citations. The source node is the solid dark grey node, pale yellow is a selected node for inspection, and green circles are referenced or citing works depending on $y$~position.

\clearpage
\bibliography{refs}

\end{document}